# Accurate Measurements of the Intrinsic Surface Impedance of Thin YBa$_2$Cu$_3$O$_{7-\delta}$ Films Using a Modified Two-tone Resonator Method

J. H. Lee, W. I. Yang, M. J. Kim, James C. Booth, K. Leong, S. Schima,

David Rudman, and Sang Young Lee

*Abstract*— We propose a modified two-tone method that could be used for sensitive measurements of the intrinsic microwave surface impedance ($Z_S$) of thin superconductor films and the *tan δ* of a low-loss dielectric. An open-gap resonator scheme is used to measure the penetration depth ($\lambda$) of thin superconductor films and extract the intrinsic $Z_S$ of the superconductor films from its measured $R_S^{eff}$ and $\lambda$. We use a very small gap of 10 μm between the top plate and the rest parts of the resonator. The *tan δ* of rutile in the low $10^{-7}$ range and the dielectric constant as high as ~110 are observed at temperatures below 10 K at ~15.2 GHz, which enable to measure the $R_S$ of the 10 mm-in-diameter YBCO films as low as ~100 μΩ at the same frequency ($f$). The discrepancy between the $R_S^{eff}$ at ~15.2 GHz and that at ~8.5 GHz scaled to ~15.2 GHz appears less than 6 % when the relations of $R_S \propto f^2$ and $tan \delta \propto f$ are used. We describe usefulness of our measurement method for measuring the intrinsic microwave properties of various superconductor samples.

Manuscript received October 5, 2004.
This work was supported in part by Korean Ministry of commerce, Industry, and Energy and Ministry of Science and Technology.

S. Y. Lee(corresponding author) is with Department of Physics and Center for Emerging Wireless Transmission Technology, Konkuk University, Seoul, Korea (phone: 82-2-450-3166; fax 82-2-2201-2759; e-mail:sylee@konkuk.ac.kr).

J. H. Lee is with Department of Physics and Center for Emerging Wireless Transmission Technology, Konkuk University, Seoul 143-701, Korea (e-mail: jaju@konkuk.ac.kr).

W. I. Yang is with Department of Physics and Center for Emerging Wireless Transmission Technology, Konkuk University, Seoul 143-701, Korea (e-mail: woori@konkuk.ac.kr).

M. J. Kim is with Department of Physics and Center for Emerging Wireless Transmission Technology, Konkuk University, Seoul 143-70l, Korea(e-mail: durance@konkuk.ac.kr).

James C. Booth is with National Institute of Standards and Technology, Boulder, CO 80305, U.S.A. (e-mail : booth@boulder.nist.gov)

K. Leong is with National Institute of Standards and Technology, Boulder, CO 80305, U.S.A. (e-mail : kleong@boulder.nist.gov)

S. Schima is with National Institute of Standards and Technology, Boulder, CO 80305, U.S.A.

David Rudman is with National Institute of Standards and Technology, Boulder, CO 80305, U.S.A. (e-mail: rudman@boulder.nist.gov).

## I. INTRODUCTION

Accurate measurements of the microwave surface impedance ($Z_S$) of superconductor films are very important for various microwave applications as well as for understanding the electrodynamics inside superconductor films. For this purpose, the dielectric resonator method has been proved very useful, with numerous studies performed by using this method [1].

When the surface resistance ($R_S$) of superconductor films is obtained from the measured quality factor of the $TE_{0mn}$ mode dielectric resonator, it is very important to know the loss tangent of the dielectric accurately. For this purpose, different methods for the *tan δ* of low-loss dielectrics have been reported [2-5].

Recently, Kobayashi *et al*. proposed the so-called 'two-tone method', which they used to measure both the *tan δ* of sapphire and the surface resistance of YBa$_2$Cu$_3$O$_{7-\delta}$ (YBCO) films simultaneously [6]. This method provides an excellent way to measure the loss tangent of the dielectric with very low loss and the surface resistance of superconductor films with accuracy. However, it is still pointed out that this method is only useful for measuring the effective surface resistance ($R_S^{eff}$) of superconductor films if the film thickness is not much larger than the penetration depth ($\lambda$). That means, it is still impossible to assess the intrinsic microwave properties of superconductor films having different thicknesses by using the 'two-tone method'.

Here we report on a modified two-tone method that enables to measure the intrinsic $R_S$ of superconductor films as well as the *tan δ* of low-loss dielectric with accuracy. We demonstrate usefulness of our modified two-tone method using a rutile-loaded resonator. For obtaining the intrinsic $R_S$, we measure $\lambda$ of the superconductor film using a ~19.6 GHz $TE_{011}$ mode sapphire resonator with a small gap between the cavity and the top plate. The measured *tan δ* values of rutile and the intrinsic $R_S$ values of the YBCO films at ~ 15.2 GHz are compared with those at ~ 8.5 GHz scaled to 15.2 GHz using the relation $R_S \propto f^2$.

## II. THEORETICAL BACKGROUND

1. The Two-tone Method

In using the two-tone method, we design a dielectric resonator having the $TE_{012}$ mode and the $TE_{021}$ mode next to each other with no parasitic mode appearing between them. The unloaded $Q$ ($Q_0$) of a dielectric-loaded resonator is described by

$$\frac{1}{Q_{0p}} = \frac{R^t_{Sp}}{G^T_p} + \frac{R^b_{Sp}}{G^B_p} + \frac{R^S_{Sp}}{G^S_p} + k_p \times tan\delta_p, \quad (1)$$

where $R^t_{Sp}$, $R^b_{Sp}$ and $R^s_{Sp}$ denote the $R_S$ values of the top plate, the bottom plate, and the side wall, respectively, $k_p$, the filling factor, $tan\delta_p$, the loss tangent of the dielectric rod, $G^T_p$, $G^B_p$, and $G^S_p$, the geometric factor of the top, the bottom and the side wall with $p=1$ for the $TE_{012}$ mode and $p=2$ for the $TE_{021}$ mode, respectively. The important part of the two tone method is that $G^T_p$, $G^B_p$, and $G^S_p$ in Eq. (1) are different between $p=1$ and $p=2$ while the $TE_{012}$ mode resonant frequency $f_1$ and that of the $TE_{021}$ mode $f_2$ are almost the same. In the two-tone method, we measure $R_S^{eff}$ of superconductor films and $tan\delta_p$ at the average frequency of $f_{av} = (f_1 + f_2)/2$. More details on the two-tone method are found elsewhere [6, 7].

2. Relationship between the Intrinsic Surface Impedance ($Z_S$) and the Effective Surface Impedance ($Z_S^{eff}$)

Analytic expressions between the $Z_S^{eff}$ and the intrinsic $Z_S$ of the superconductor films are obtained by using a $TE_{0mn}$ mode field analysis based on a mode matching method. Although descriptions below are strictly based on the $TE_{0mn}$ mode of the dielectric resonator, we note that the obtained results for the $Z_S^{eff}$ vs. $Z_S$ relation applies to the other $TE_{0mn}$ modes.

For a $TE_{0mn}$ mode dielectric resonator as shown in Fig. 1, the field distributions in the $k$-th region are given by $H_{zk} = A_k$ x $q_k(\beta_{zk}z)$ x $\Psi_k(r)$, $E_{\varphi k} = (j\omega\mu_0/\beta_k^2)$ x $A_k$ x $q_k(\beta_{zk}z)$ x $d\Psi_k/dr$, and $H_{rk} = A_k$ x $dq_k(\beta_{zk}z)/dz$ x $d\Psi_k/dr$ with $E_{rk} = H_{\varphi k} = 0$ and $k = 1 – 5$. In expressing the field components, the origin is set at the center of the dielectric and $j^2 = -1$. According to Zaki et al. [8], the non-zero field components for regions 1 and 2 are given as follows.

$$H_{z1} = A\, J_0(\beta_1 r)\, q_1(\beta_{z1}z),$$
$$H_{r1} = (A/\beta_1)\, J_0'(\beta_1 r)\, dq_1(\beta_{z1}z)/dz, \quad (2)$$
$$E_{\varphi 1} = (j\omega\mu_0 A/\beta_1)\, J_0'(\beta_1 r)\, q_1(\beta_{z1}z),$$

for region 1 and,

$$H_{z2} = A\, Q_0(\beta_1 r)\, q_2(\beta_{z2}z),$$
$$H_{r2} = (A/\beta_2)\, Q_0'(\beta_1 r)\, dq_2(\beta_{z2}z)/dz, \quad (3)$$
$$E_{\varphi 2} = (j\omega\mu_0 A/\beta_1)\, Q_0'(\beta_1 r)\, q_2(\beta_{z2}z),$$

for region 2.

Here $J_0(x)$ denotes the zeroth order Bessel function of the first

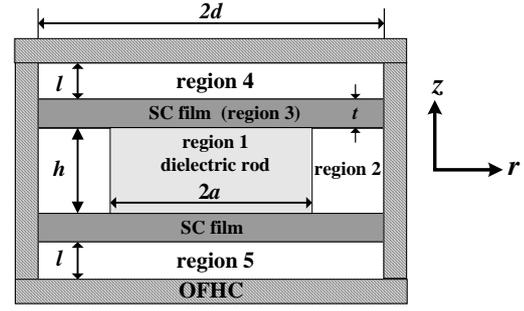

Fig. 1. A diagram of for the cross-sectional view of our dielectric-loaded resonator. Here region 2 is vacuum, region 3 is for superconductor (SC) film, and region 4(5) are for the dielectric substrate. The regions 2, 4 and 5 are surrounded by OFHC.

For field analysis, we set the origin is at the center of the dielectric.

kind, $q_1(\beta_{z1}z) = q_2(\beta_{z2}z) = cos(\beta_{z1}z + \psi_h)$ with $\beta_{z1} = \beta_{z2}$,

$$Q_0(\beta_2 r) = \frac{-\beta_2 J_0'(\beta_2 a)}{\beta_1} \cdot \frac{K_0(\beta_2 r)I_0'(\beta_2 d) - I_0(\beta_2 r)K_0'(\beta_2 d)}{K_0(\beta_2 a)I_0'(\beta_2 d) - I_0(\beta_2 a)K_0'(\beta_2 d)}, \quad (4)$$

and $A$ denoting a constant.

Also, for $k = 1, 2$, $\beta_k$ and $\beta_{zk}$ are related by $k_0^2\varepsilon_{r1} = \beta_{z1}^2 + \beta_1^2$ and $k_0^2\varepsilon_{r2} = \beta_{z2}^2 - \beta_2^2$. Here $\varepsilon_{r1}$ and $\varepsilon_{r2}$ denote the dielectric constant of the dielectric rod and the region 2, respectively, $k_0 = \omega_0\sqrt{\varepsilon_0\mu_0}$, $\omega_0 = 2\pi f_0$ with $f_0$ denoting the resonant frequency, and $\beta_1$ and $\beta_2$, the propagation constants along the transverse direction in regions 1 and 2, respectively. Also the boundary condition of $E_{\varphi 1}/H_{z1} = E_{\varphi 2}/H_{z2}$ at $r = a$ gives

$$\frac{\beta_2 J_1(\beta_1 a)}{\beta_1 J_0(\beta_1 a)} = \frac{I_1(\beta_2 a)K_1(\beta_2 d) - I_1(\beta_2 d)K_1(\beta_2 a)}{I_1(\beta_2 d)K_0(\beta_2 a) + I_0(\beta_2 a)K(\beta_2 d)} \quad (5)$$

for $k_0^2\varepsilon_2 < \beta_{z2}^2$ for the $TE_{0mn}$ mode. In obtaining Eq. (5), we used the relations of $J_0'(x) = -J_1(x)$, $K_0'(x) = -K_0(x)$ and $I_0'(x) = I_0(x)$. If $k_0^2\varepsilon_2 > \beta_{z2}^2$, the modified Bessel functions $I_n$ and $K_n$ in $Q_0(\beta_2 r)$ and Eq. (5) are replaced by the corresponding Bessel functions $J_n$ and $Y_n$, respectively.

For $k = 4$ and 5, with the two superconductor films in region 3 assumed identical, $q_k(\beta_{zk}z) = exp(-i\beta_{zk}z) - \Gamma_k exp(i\beta_{zk}z)$ with $\beta_{z4} = -\beta_{z5}$ if traveling modes exists in regions 4 and 5 with $q'_k(\beta_{zk}z) = dq_k(\beta_{zk}z)/d(\beta_{zk}z)$ and $\Psi_k(r) = J_0(\beta_k r)$ for $k = 4$ and 5. In this case, $\beta_4$ and $\beta_{z4}$ are related by $\beta_{z4}^2 = k_0^2\varepsilon_{r4} - \beta_4^2$ with $\varepsilon_{r4}$ denoting the dielectric constant in regions 4 and 5, respectively. If $\beta_4^2 > k_0^2\varepsilon_{r4}$, then $\beta_{z4}$ is replaced by $\gamma_{z4}$ with $\gamma_{z4} = j\beta_{z4}$ to account for the existence of evanescent modes in regions 4 and 5.

In region 3, $q_3(\gamma_{z3}z)$ and $q_3'(\gamma_{zk}z)$ can be expressed by $q_3(\gamma_{z3}z) = sinh(\gamma_{z3}z) + \Gamma_3 cosh(\gamma_{z3}z)$ and $q_3'(\gamma_{z3}z) = cosh(\gamma_{z3}z) + \Gamma_3 sinh(\gamma_{z3}z)$. It is noted that $H_{z3} = 0$ in region 3 if the magnitude of the magnetic field is less than the lower critical field of the superconductor $H_{C1}$, and $\gamma_{z3} = (j\omega\mu_0\sigma)^{1/2}$ with $\mu_0$, the permeability of the vacuum and $\sigma$ ($=\sigma_1 - i\sigma_2$), the complex conductivity of the superconductor in the region.

From the boundary conditions of $E_{\varphi 4} = 0$ at $r = d$ and $E_{\varphi 4} = 0$ at $z = l + h/2 + t$ in region 4, we get $\beta_4 = \mu_{0m}/d$ with $\mu_{0m}$ denoting the *m*-th roots of the derivative of the Bessel functions $J'_0(x)$ and $\Gamma_4 = exp[-\beta_{z4}(2t + 2l + h)]$, respectively [9]. From another boundary condition of $E_{\varphi3}/H_{r3} = E_{\varphi4}/H_{r4}$ at $z = h/2 + t$ between regions 3 and 4, $\Gamma_3$ is expressed by

$$\Gamma_3 = \frac{(\beta_h/\gamma_{z3}) \sinh(\gamma_{z3}(h/2+t)) - \cosh(\gamma_{z3}(h/2+t))}{\sinh(\gamma_{z3}(h/2+t)) - (\beta_h/\gamma_{z3})\cosh(\gamma_{z3}(h/2+t))} \quad (6)$$

with $\beta_h = -\beta_{z4} \cot(\beta_{z4} l)$. Also, from the boundary condition of $H_{r1}/E_{\varphi1} = H_{r3}/E_{\varphi3}$ at $z = h/2$ and Eq. (6) for $\Gamma_3$, we get $\beta_{z1}\tan(\beta_{z1}h/2 + \psi_h) = \gamma_{z3}/G_h^*$ with $G_h^*$ given by

$$G_h^* = \frac{\coth(\gamma_{z3}t) - (\beta_h/\gamma_{z3})}{1 - (\beta_h/\gamma_{z3})\coth(\gamma_{z3}t)}. \quad (7)$$

Finally, for the $TE_{0mn}$ modes, $q_k(-\beta_{zk}z) = q_k(\beta_{zk}z)$ if *n* is odd and $q_k(-\beta_{zk}z) = -q_k(\beta_{zk}z)$ if *n* is even, for $k = 1, 2$. In this case, $\psi_h = s\pi$ for odd *n* and $\psi_h = (2s + 1)\pi/2$ with *s* denoting an integer and we get

$$\beta_{z1}\tan(\beta_{z1}h/2) = \gamma_{z3}/G_h^* \text{ (odd } n\text{), or}$$
$$-\beta_{z1}\cot(\beta_{z1}h/2) = \gamma_{z3}/G_h^* \text{ (even } n\text{)} \quad (8)$$

At temperatures not too close to the critical temperature ($T_C$), the resonant frequency of the resonator can be calculated from Eqs. (4) - (8), the relations of $\beta_{z1} = \beta_{z2}$, $\beta_4 = \mu_{0m}/d$, the equations for $\beta_k$'s and $\beta_{zk}$'s and $\gamma_{z3} \approx 1/\lambda$ if $\lambda$ is known.

$Z_S^{eff}$ of the superconductor films can be obtained from the ratio of $E_{\varphi3}$ to $H_{r3}$, such that

$$Z_S^{eff} = -\frac{E_{\varphi3}}{H_{r3}}|_{z=h/2} = \frac{i\omega\mu_0}{\gamma_{z3}} G_h^* = Z_S \frac{\beta_h - \gamma_{z3}\coth(\gamma_{z3}t)}{\beta_h\coth(\gamma_{z3}t) - \gamma_{z3}}. \quad (9)$$

The expression for $Z_S^{eff}$ from our TE mode analysis is similar to that obtainable by two successive impedance transformations at the surfaces of the substrate and the superconductor film [10]. That is, $Z_S^{eff}$ can also be rewritten by $Z_S^{eff} = \{\eta_4^{eff} + Z_S \tanh(\gamma_{z3}t)\}$ x $Z_S/\{Z_S + \eta_4^{eff} \tanh(\gamma_{z3}t)\}$ with $\eta_4^{eff} = -\frac{E_{\varphi4}}{H_{r4}}|_{z=t+h/2}$ denoting the effective surface impedance of the substrate. We note, however, the $\eta_4^{eff}$ value used in the impedance transformation method is different from that in the TE mode analysis due to the difference in $\beta_{z4}$ in the two methods: $\beta_{z4}^2 = k_0^2\varepsilon_4 - \beta_4^2$ in the TE mode analysis while $\beta_{z4}^2 = k_0^2\varepsilon_4$ in the impedance transformation method.

3. Measurements of the Penetration Depth

If the shift in the resonant frequency of a dielectric resonator is solely attributed to the changes in the surface reactance of the YBCO film at the top of the cavity, the variation in the effective penetration depth $\Delta\lambda_{eff}$ of the top YBCO film is given by $\Delta\lambda_{eff} = -(\Delta f_0(T)/f_0(T)^2)/(\pi\mu_0 B_i)$ with $\Delta f_0(T) = f_0(T) - f_0(T_{min})$, $T_{min}$ and $f_0$ denoting the measured minimum temperature and the relevant mode resonant frequency, $\mu_0$, the magnetic permeability of free space, and $B_i$, the inverse of the geometric factor for the top plate corresponding to the resonant mode, respectively. $\Delta\lambda_{eff} \sim$ $K\Delta\lambda$ at $T < (2/3) T_C$ with *K* denoting the real part of the ratio of the effective surface impedance $Z_S^{eff}$ to the intrinsic surface impedance $Z_S$ (i.e., $K = Re(Z_S^{eff}/Z_S)$), and the fitted value of $\lambda$ can be obtained from a fit to $\Delta\lambda_{eff}$ using $\Delta\lambda_{eff} = K\Delta\lambda$ and the model equation for $\lambda$ of the superconductor films of interest. If the resonant frequency of the resonator is not close to that of the substrate, the description for *K* is almost the same as that obtained by impedance transformation method [10]. We note that, in general, the numerical solutions for the $\sigma_1$ and $\sigma_2$ values are obtained from Eq. (9) and $\gamma_{z3} = (j\omega\mu_0\sigma)^{1/2}$ using the measured $R_S^{eff}$ and $\Delta\lambda_{eff}$. Then, the intrinsic $R_S$ is obtained from $Re(Z_S) = Re\{(j\mu_0\omega_0/\sigma)^{1/2}\}$. As the model equation for the temperature-dependent $\lambda$ of YBCO films, we use $\lambda = \lambda_0[1-(T/T_C)^2]^{-1/2}$.

III. EXPERIMENTAL

Two 300 nm-thick YBCO films with the critical temperature of ~87 K were prepared on round LaAlO$_3$ (001) substrates with the diameter of 10 mm under the same preparation conditions. A rutile-loaded cavity was fabricated using a rutile rod having the dimensions of 3.9 mm in diameter and 2.2 mm in height, with the YBCO films used as the endplates. The cavity is made of oxygen-free high-purity copper (OFHC) with the diameter of 9 mm. We note that rutile is well known for its very small $\tan\delta$ (e.g., ~10$^{-6}$ at 77 K) despite that the dielectric constant is very high (e.g., ~110 at 77 K). The resonator is weakly coupled throughout the measurements and the loaded *Q* ($Q_L$) values are obtained with the insertion loss more than 30 dB throughout the measurements. Loop coupling is used for the TE mode excitations. Cares needs to be taken to realize symmetric coupling to the resonator, which enables to get $Q_0$ from $Q_0 = Q_L/(1 - 10^{-IL(dB)/20})$ with $IL(dB)$ denoting the insertion loss in decibel. The $Q_0$ values obtained using the *S*-parameter circle-fit method [11] appear almost the same as that by the insertion-loss method. More descriptions on the rutile-loaded cavity are reported elsewhere [12]. For obtaining the $R_S^{eff}$ of the YBCO film at the top (i.e., $R_{Sp}^T$ in Eq. (1)) and $\tan\delta$ of rutile, we used $G_1^T = G_1^B = 171.6$ Ω, $G_1^S = 3.57195$ x $10^8$ Ω, $k_1 = 0.999617$ at 15.099 GHz for the $TE_{012}$ mode and $G_2^T = G_2^B = 572$ Ω, $G_2^S = 294247$ Ω, and $k_2 = 0.997459$ at 15.253 GHz for the $TE_{021}$ mode at 10 K, respectively. The $R_{S1}^S$ and $R_{S2}^S$ values of OFHC were separately measured at the $TE_{012}$ and the $TE_{021}$ mode frequencies using the rutile-loaded resonator with the top and the bottom plates made of OFHC. $\lambda$ of the YBCO films could be measured using a ~19.6 GHz $TE_{011}$ mode sapphire resonator with a 10 μm gap between the top and the rest parts of the resonator, as shown in Fig. 2. The dimensions of the sapphire-loaded cavity are 9 mm and 5 mm for the diameter and height, respectively, with the dimensions of sapphire being 5 mm in diameter and 5 mm in height. The $TE_{011}$ mode resonant frequency is ~19 GHz for the open-gap resonator. In this scheme, the gap between the cavity and the top plate enables to control the temperature of the top YBCO film separately while maintaining the temperature of the cavity at the same temperature within ± 0.5 K. That means, the shift in the mode frequency of the resonator is solely attributed to the changes in the surface reactance of the YBCO film. We note that the existence of the extremely small gap affects the shift in the resonant frequency within 5 %. $\lambda$ at 15.2 GHz is assumed to be equal to that at ~19 GHz due to the frequency-independent nature of $\lambda$ except at temperatures very close to the critical temperature. Later we compare the measured values for the intrinsic $R_S$ and $\lambda$ at 15.2 GHz with those at ~8.5 GHz for which the $\tan\delta$ values of rutile are measured by using the reported $\tan\delta$ of sapphire as the reference values [4]. We note that the reason for using an open-gap sapphire resonator separately for $\lambda$ of the YBCO films is because the dielectric constant of rutile changes significantly even for a temperature fluctuation of ± 0.5 K. If a sapphire resonator is prepared for the two-tone measurement, the same resonator is better to be used for the $\lambda$ measurements.

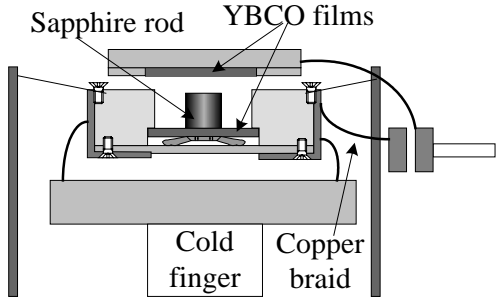

Fig. 2. A schematic diagram for the open-gap rutile resonator installed inside the cryostat. The thermal path to the top plate from the cold finger is designed to be disconnected from outside for independent temperature control of the top plate while the other part of the resonator remains at the same temperature.

## IV. RESULTS AND DISCUSSION

The dimensions of the rutile and the cavity are determined such that the two modes appear next to each other without being coupled to any other parasitic modes. In Fig. 3, the simulated results obtained by 'Microwave Studio' are seen for the rutile resonator [13], where the two modes are separated by ~113 MHz with the resonant frequency $f_1$ ~ 15.160 GHz for the $TE_{012}$ mode and the resonant frequency $f_2$ ~ 15.274 GHz for the $TE_{021}$ mode. For the simulation, we use ~108 for the dielectric constant of rutile, a value measured at ~77 K. These values are close to the calculated ones from the analytic expressions described in II.2, which are 15.177 GHz and 15.291 GHz for the $TE_{012}$ and the $TE_{021}$ modes, respectively. The nearest two parasitic modes above the $TE_{021}$ and below the $TE_{012}$ modes are seen at ~14.964 GHz ($HEM_{120}$) and ~15.593 GHz ($HEM_{021}$) in the figure.

The measured values for the resonant frequency of the two modes appear very close to the simulated and the calculated ones from field analysis. Figure 3 also show the frequency response of the resonator measured at 10 K, where the two modes are seen at 15.099 GHz ($TE_{012}$) and 15.253 GHz ($TE_{021}$). In Table 1, we show a list of values for the measured $Q_0$, the resonant frequency, the geometric factor and the filling factor for the $TE_{011}$, $TE_{012}$ and $TE_{021}$ modes.

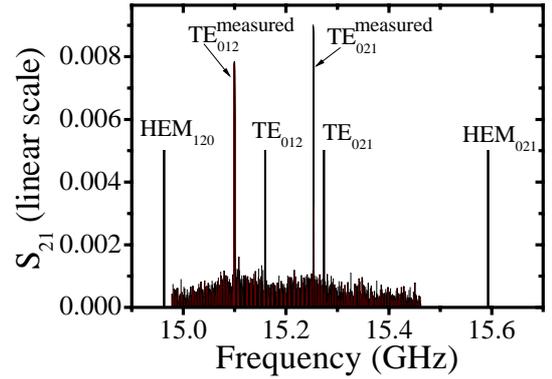

Fig. 3. The frequency response of our rutile resonator compared with the simulated results. The two HEM modes close to the $TE_{012}$ and $TE_{021}$ modes seen in the simulated data cannot be identified in the measured one.

We display the temperature dependence of the $Q_0$ values of the two modes in Fig. 4 at temperatures above 6 K, where $Q_0$ appear to change continuously as the temperature changes. This is in contrast with what we observed from another rutile-loaded resonator over a certain temperature range [7], signaling that there is no mode coupling between the two modes and the nearby parasitic modes throughout the temperature range. In Fig. 4, the $TE_{012}$ mode and $TE_{021}$ mode $Q_0$ values are ~500000 and ~1200000 at ~ 6 K. The inset of Fig. 4 shows the temperature dependence of the resonant frequency for the two modes, where the increase of the resonant frequency for the two modes is due to the decrease of the dielectric constant of the rutile for increasing temperature. For obtaining the $R_S$ of the superconductor films, we measured the $R_S$ of OFHC separately at the frequencies $f_1$ and $f_2$. For the measurement, we replaced the superconducting endplates by thick OFHC plates and obtained the $R_S$ values of OFHC from the measured $Q_0$ of the two modes (Note that these values are the $R_{Sp}$ in Eq. (1)). We used the separately measured values for the $R_S$ of OFHC at $f_1$ and $f_2$ instead of using the relation of $R_S$ ~ $f^{1/2}$ due to that copper shows the anomalous skin effect at low temperatures.

TABLE I.
LIST OF VALUES FOR THE MEASURED $Q_0$, THE RESONANT FREQUENCY, THE GEOMETRIC FACTOR AND THE FILLING FACTOR FOR THE $TE_{011}$, $TE_{012}$ AND $TE_{021}$ MODES AT 10 K.

|  | $TE_{011}$ | $TE_{012}$ | $TE_{021}$ |
| --- | --- | --- | --- |
| $Q_0$ | 932000 | 439000 | 1048500 |
| $f_0$ (GHz) | 9.5541 | 15.0990 | 15.253 |
| $G^T$ (=$G^B$) ($\Omega$) | 157 | 172 | 572 |
| $G^S$ ($\Omega$) | 131965 | 3.57 x $10^8$ | 294247 |
| $k$ | 0.9985 | 0.9996 | 0.9975 |

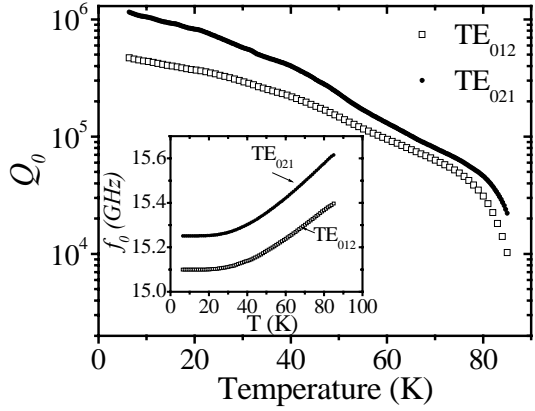

Fig. 4. The temperature dependence of the TE$_{012}$ and TE$_{021}$ mode $Q_0$ of the rutile resonator. Inset: The temperature dependence of the resonant frequency of the TE$_{012}$ and TE$_{021}$ modes.

The temperature dependence of the $R_S^{eff}$ of the YBCO films at $f_{av} = (f_1 + f_2)/2$ measured by the two-tone method is seen in Fig. 5, with its inset showing the corresponding $tan\delta$ values of rutile. We get the $tan\delta$ values at $f_1$ and $f_2$ from those at $f_{av}$, assuming the relation of $tan\delta \propto f$, which enable to obtain the respective $R_S^{eff}$ values at $f_1$ and $f_2$ from Eq. (1). For obtaining the $R_S^{eff}$ values as a function of the temperature, we considered the temperature dependence of $G^T_p$ ($=G^B_p$), $G^S_p$ and $k$. The $tan\delta$ values are ~$3.5\times10^{-7}$ and ~$1.0 \times 10^{-5}$ at 6 K and 77 K, respectively, at ~15.17 GHz. The $R_S^{eff} \propto f^2$ relation appeared to hold between the two sets of values at $f_1$ and $f_2$ within 2 %, demonstrating the accuracy of the measured $R_S^{eff}$ values. Furthermore the $R_S^{eff}$ values at $f_1$ appear to match well with those at ~8.5 GHz scaled to $f_1$, as seen in Fig. 5. The data at ~8.5 GHz are collected using a separate TE$_{011}$ mode rutile resonator, for which the $tan\delta$ values of the rutile scaled to 15.17 GHz are seen in the inset of Fig. 5.

We note in the inset of Fig. 5 that the $tan\delta$ values appear significantly different between the two at temperatures below 30 K, with the $tan\delta$ measured at ~15.17 GHz becoming almost twice larger than that at ~8.5 GHz scaled to ~15.17 GHz. Considering the $tan\delta$ of low-loss dielectrics strongly dependent on the impurity level, the observed difference in the $tan\delta$ at low temperatures is likely due to the difference in the crystal quality

It is seen in the figure that the $R_S^{eff}$ of the YBCO films about ~ 100 μΩ can be safely measured at low temperatures as far as $R_{Sp}^T / G^T_p \sim k_p \times tan\delta_p$ in Eq. (1). This means that the intrinsic $R_S$ of thin superconductor films on the order of 10 μΩ can be measured with accuracy after the thickness correction, with even better sensitivity achievable at lower frequencies with high-quality rutile having lower $tan\delta$. In this context, our data show that the impurity effect becomes stronger at low temperatures.

The temperature-dependent resonant frequency of the open-ended $TE_{011}$ mode sapphire resonator is seen in Fig. 6, with the inset displaying the variations in $\Delta\lambda_{eff}$ at different temperatures. Good agreements are seen at temperatures below 80 K with the best-fitted value of $\lambda_0 = 185$ nm for the YBCO film. This is comparable to the value of 150 – 200 nm, that have been reported for $\lambda$ of epitaxially grown YBCO films. Differences between $\Delta\lambda_{eff}$(exp) and $\Delta\lambda_{eff}$(fitted) at T > 80 K are likely due to the failure of the relation $\Delta\lambda_{eff} = K\Delta\lambda$ and $\lambda = \lambda_0[1-(T/T_C)^2]^{-1/2}$ near $T_C$. Further studies are under progress to clarify this. After obtaining $\lambda$, we proceed to get the intrinsic $R_S$ of the YBCO film. In Fig. 7, we show the temperature dependence of the $R_S$ at ~8.5 GHz and ~15.17 GHz with the inset showing the $R_S$ at ~15.17 GHz compared with that of $R_S$ at ~8.5 GHz scaled to ~15.17 GHz using the $R_S \propto f^2$ relation. We see in the figure that the two sets of the $R_S$ values agree well with each other with the differences of less than ~6 % at temperatures below 86 K.

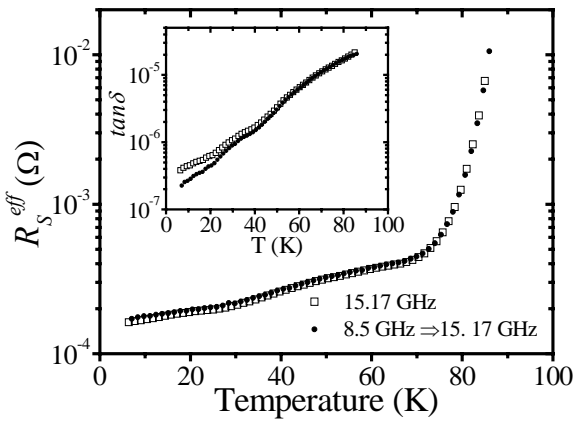

Fig. 5. Comparison between $R_S^{eff}$ of a YBCO film measured at ~15.17 GHz by the two-tone method and that at ~8.5 GHz scaled to ~15.17 GHz. The two sets of data agree well within 6 % at temperatures below 86 K. Inset: The measured $tan\delta$ of rutile at ~15.17 GHz used for the two-tone method.

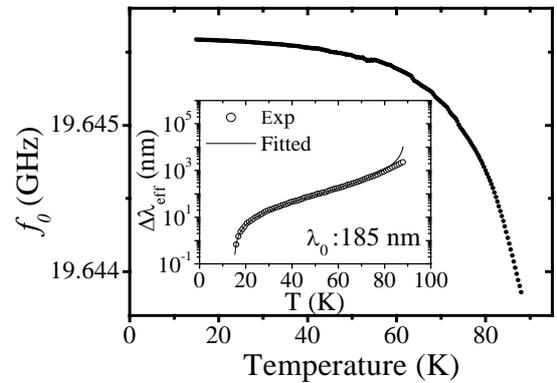

Fig. 6. The temperature dependence of the ~19.6 GHz $TE_{011}$ mode open-gap sapphire resonator. Inset: Variations in the effective penetration depth on the temperature for the YBCO film used as the top plate of the resonator. Good agreements are seen between the measured values and the fitted ones with $\lambda_0 = 185$ nm.

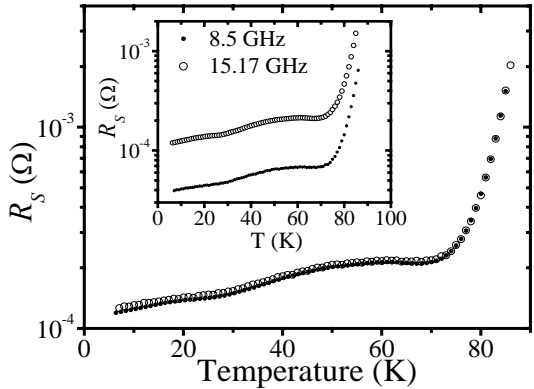

Fig. 7. The temperature dependence of the intrinsic $R_S$ of the YBCO film. The $R_S$ at ~15.17 GHz (open) appear to agree well with those at ~8.5 GHz scaled to ~15.17 GHz (dot) using the $R_S \propto f^2$ relation within 6 % at temperatures below 86 K. Inset: The $R_S$ of the YBCO film measured at ~15.17 GHz (open) and ~8.5 GHz (dot).

Our results show that even the $tan\delta$ of rutile having a strong temperature dependence can be measured with accuracy by using a resonator properly designed for the two-tone method. The relatively low $tan\delta$ of rutile combined with the high dielectric constant would make it possible to measure the microwave properties of superconductor samples which is too small to be measured with sapphire resonator due to the limits in the measurement frequency.

## V. CONCLUSIONS

A modified two-tone measurement method is proposed for sensitive measurements of the intrinsic $R_S$ and $\lambda$ of the YBCO films as well as $tan\delta$ of low-loss dielectric. A rutile-loaded resonator with YBCO films used as the endplates is designed and tested, where the two relevant resonant modes (i.e., the $TE_{012}$ and $TE_{021}$ modes) appear close to each other enabling to measure $R_S^{eff}$ of the YBCO films and the $tan\delta$ of rutile at ~15.17 Hz at temperatures of 6 - 90 K. The dimensions of the rutile are 3.9 mm in diameter and 2.2 mm in height.

At temperatures not too close to $T_C$, the relation of $R_S^{eff} \propto f^2$ appears to hold within the discrepancy of less than 2 % between the $R_S^{eff}$ values measured at the resonant frequency of the $TE_{012}$ and $TE_{021}$ modes. The $tan\delta$ values appear significantly dependent on the crystal quality of rutile at temperatures below 30 K, with significantly different $tan\delta$ values observed for different rutile rods.

Accurate measurements of $tan\delta$ of low-loss dielectric and the penetration depth of superconductor films using an open-gap dielectric resonator scheme would enable to characterize the intrinsic microwave properties of various superconductor samples in a non-invasive way.

## ACKNOWLEDGMENTS

SYL deeply thanks the late V. B. Fedorov who initiated parts of this work during his stay at Konkuk University and Jung Hur for useful discussions. Parts of this work by SYL were done during his stay at NIST for sabbatical leave.